\documentclass[20pt,prb,twocolumn,aps,showpacs,fixfloats]{revtex4}
\usepackage{graphicx}
\usepackage{bm}
\usepackage{amsmath,amssymb}
\usepackage{subfigure}
\usepackage{float}
\usepackage{latexsym}
\usepackage{color}
\usepackage{enumerate}
\usepackage{pdfpages}
\usepackage{tikz}
\usepackage{hyperref}
\usepackage{mathtools}

\begin{document}
\newcommand{\s}{\scriptscriptstyle}
\newcommand{\uu}{\uparrow \uparrow}
\newcommand{\ud}{\uparrow \downarrow}
\newcommand{\du}{\downarrow \uparrow}
\newcommand{\dd}{\downarrow \downarrow}
\newcommand{\ket}[1] { \left|{#1}\right> }
\newcommand{\bra}[1] { \left<{#1}\right| }
\newcommand{\bracket}[2] {\left< \left. {#1} \right| {#2} \right>}
\newcommand{\vc}[1] {\ensuremath {\bm {#1}}}
\newcommand{\tr}{\text{Tr}}
\newcommand{\Trans}{\ensuremath \Upsilon}
\newcommand{\Refl}{\ensuremath \mathcal{R}}

\title{Enhanced interaction effects in the vicinity of the topological transition}

\author{C. C. A. Houghton, E. G. Mishchenko,   and M. E. Raikh}

\affiliation{ Department of Physics and
Astronomy, University of Utah, Salt Lake City, UT 84112}
\begin{abstract}
A metal near the topological transition can be loosely viewed as consisting of two groups
of electrons. First group are ``bulk" electrons occupying most of the Brillouin zone.
Second group are electrons with wave vectors close to the topological transition point.
Kinetic energy, ${\tilde E}_{\s F}$, of electrons of the first group is much bigger
than kinetic energy, $E_{\s F}$, of electrons of the second group. With electrons
of the second group being slow, the interaction effects  are more pronounced for
these electrons. We perform a calculation illustrating that electrons of the second
group are responsible for inelastic lifetime making it anomalously short,
so that the concept of quasiparticles applies to these electrons only marginally.
We also demonstrate that interactions renormalize the spectrum of electrons in the vicinity
of topological transition, the parameters of renormalized spectrum being strongly dependent
on the proximity to the transition. Another many-body effect that evolves dramatically
as the Fermi level is swept through the transition is the
Friedel oscillations of the electron
density created by electrons of the second group around an impurity.
These oscillations are strongly anisotropic
with a period depending on the direction. Scattering of electrons off these oscillations give
rise to a temperature-dependent ballistic correction to the conductivity.

\end{abstract}

\maketitle

\section{Introduction}

\begin{figure}[h!]
\includegraphics[scale=0.5]{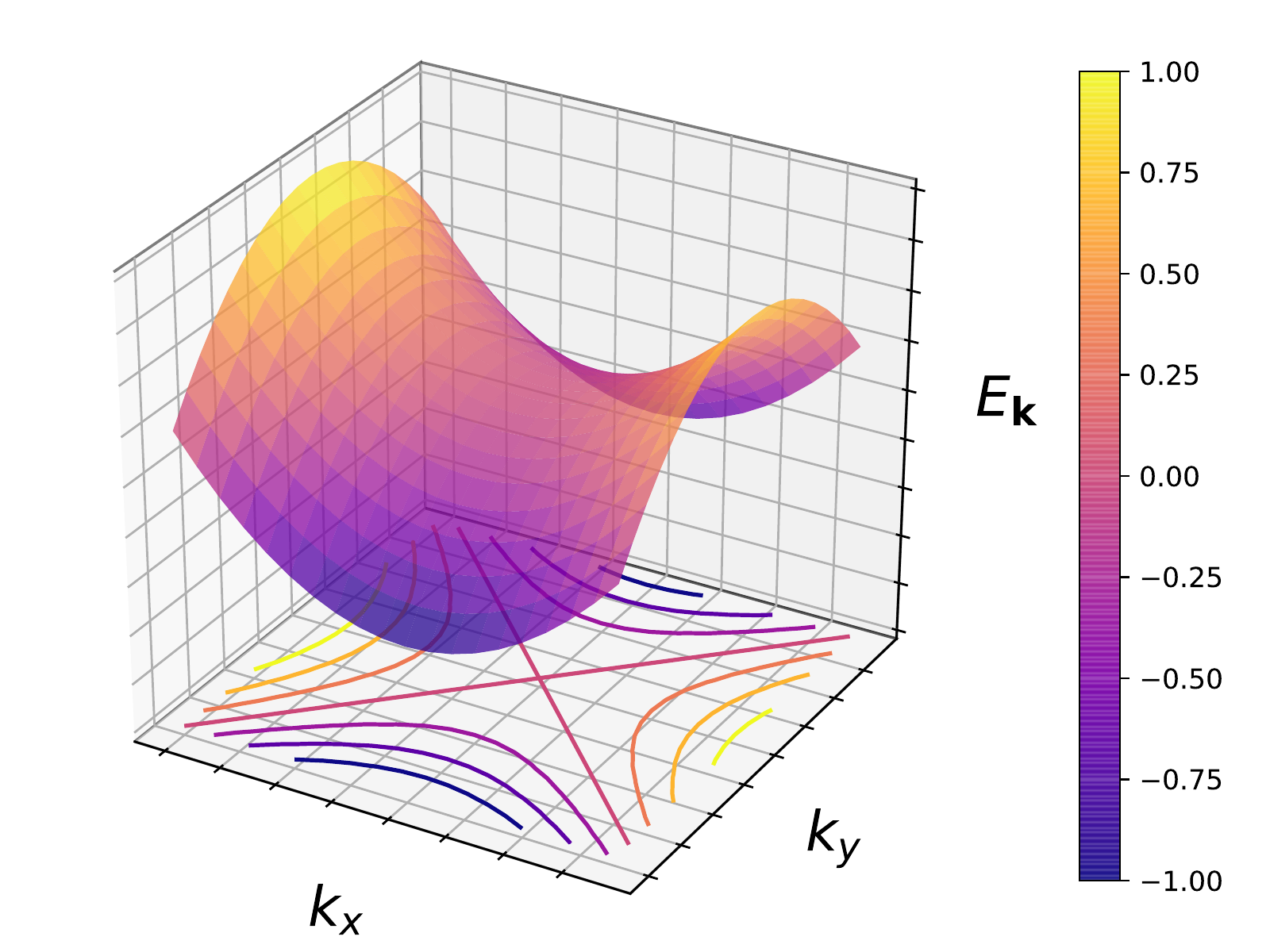}
\caption{(Color online) Electron spectrum near the
topological transition is plotted from Eq. (\ref{hyperbolic}) together with
Fermi contours for different values of the Fermi energy, $E_{\s F}$.     }
\label{f1}
\end{figure}

Topological transitions in metals take place
when, upon the change of a certain external parameter,
the connectivity
of the Fermi surface undergoes a transformation. The concept of topological
transition was introduced by I. M. Lifshitz in
1960.\cite{Lifshitz1960}. Lifshitz demonstrated that thermodynamic characteristics of a
metal exhibit a singular behavior in the vicinity of the transition. Such a singular behavior was
subsequently observed experimentally \cite{Experiment1983,Experiment1984}.
First experiments were conducted on 3D metallic alloys\cite{Experiment1983} and 2D semiconductor
superlattices.\cite{Experiment1984}
In the past decade the class of materials in which the signatures of the topological transitions were uncovered
has significantly broadened\cite{1,2,3,4,5,6,7,8} to include heavy fermions, graphite, germanene, cilicene, ruthinades, etc.

On the theoretical side,  kinetic and thermodynamic characteristics of metals near the  topological transitions\cite{Varlamov1985,Varlamov1986,Blanter1990,Blanter1991,Ablyazov1991,Golosov1991,Mobius} were actively studied
after the experiments.\cite{Experiment1983,Experiment1984} The results
are reviewed in Ref. \onlinecite{VarlamovReview}. On the conceptual level, the main  theoretical finding is that, in addition to the single-particle density of states, the transition manifests itself in the energy dependence of the impurity scattering time of carriers, which, at the same time, broaden the transition.
Recent theoretical interest to the topological transitions\cite{Weyl,Thorus,Chi-Ken2016,Graphene,Volovik,Galperin} is mostly motivated by the invent of new materials with strong spin-orbit coupling.

 The role played by electron-electron interactions in the topological transition was considered in Ref. \onlinecite{Mobius}
with a general conclusion that, away from Pomeranchuk instability, Fermi-liquid effects
renormalize the singular part of thermodynamic quantities.

The goal of the present paper is to trace how the standard many-body effects for an isotropic spectrum
get modified in the vicinity of the topological transition. We will consider the following effects:
Friedel oscillations of the electron density, interaction-induced modification of the
electron spectrum, and the interaction-induced electron lifetime caused by the creation of the electron-hole pairs.
 We find that the proximity to the transition gives rise to additional Friedel oscillations with
very long period, which are strongly anisotropic and get rotated by $90^{\circ}$ as the Fermi level is swept through
the transition.

Our main finding is that electron lifetime, $\tau_e$, associated with creation of pairs,  shortens dramatically  in the vicinity of the transition. Directly at the transition, we have $\frac{\hbar}{\tau_e(E)}\sim E$,
so that the concept of the Fermi liquid applies only marginally.

\begin{figure}[h!]
\includegraphics[scale=0.59]{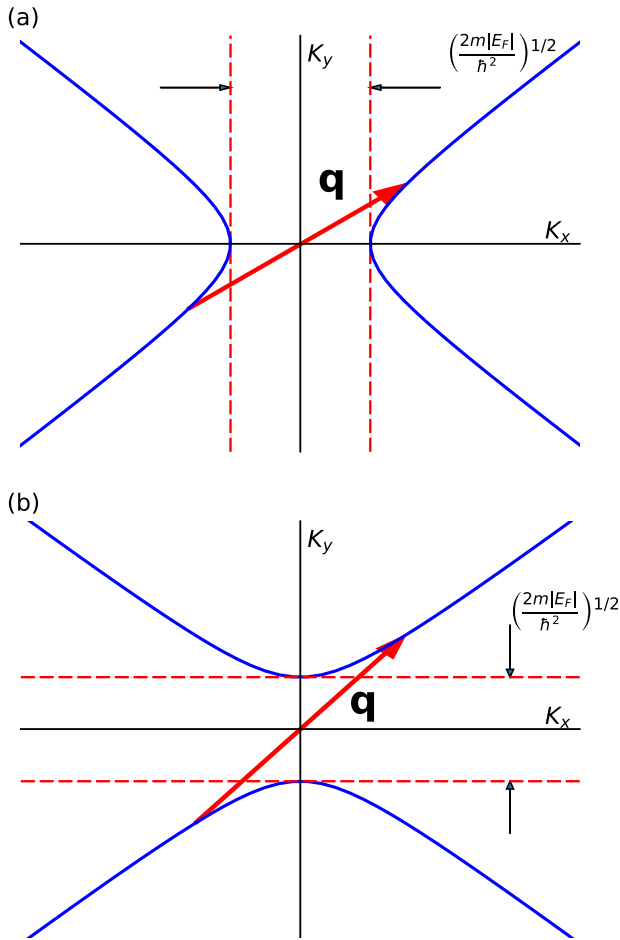}
\caption{(Color online) The process responsible for the formation of
the Kohn anomaly and ensuing long-period Friedel oscillations is illustrated
for positive $E_{\s F}$ (a) and negative $E_{\s F}$ (b). The components of the
vector ${\bf q}$ satisfy the condition $q_x^2-q_y^2=4k_{\s F}^2$, Eq. (\ref{Kohn}).}
\label{f2}
\end{figure}

Since the interaction effects are more pronounced in two dimensions,
we will choose the simplest form of the spectrum in the vicinity of the topological transition
\begin{equation}
\label{hyperbolic}
E_{\bf k} =\frac{\hbar^2k_x^2}{2m}-\frac{\hbar^2k_y^2}{2m},
\end{equation}
see Fig. \ref{f1}. The transition corresponds to the position of the Fermi level $E_F=0$.
As $E_F$ changes from negative to positive, the Fermi surface near $k_x=k_y=0$
evolves as illustrated in Fig. \ref{f2}.
Most importantly, the typical wave vector,  $\left(\frac{mE_F}{\hbar^2}\right)^{1/2}$,
in the vicinity of the transition is small, while everywhere else in the
Brillouin zone this wave vector is big, namely it is of the order of
$\left(\frac{m{\tilde E}_F}{\hbar^2}\right)^{1/2}$,
where ${\tilde E}_F$ is the Fermi level measured from the bottom of the band.

While within a single-particle
approach the topological transition causes a singular correction to the electron
characteristics, interaction effects give rise to new {\em distinct} features,
in particular, new Friedel oscillations  and a new channel
of inelastic relaxation.

\begin{figure}[h!]
\includegraphics[scale=0.55]{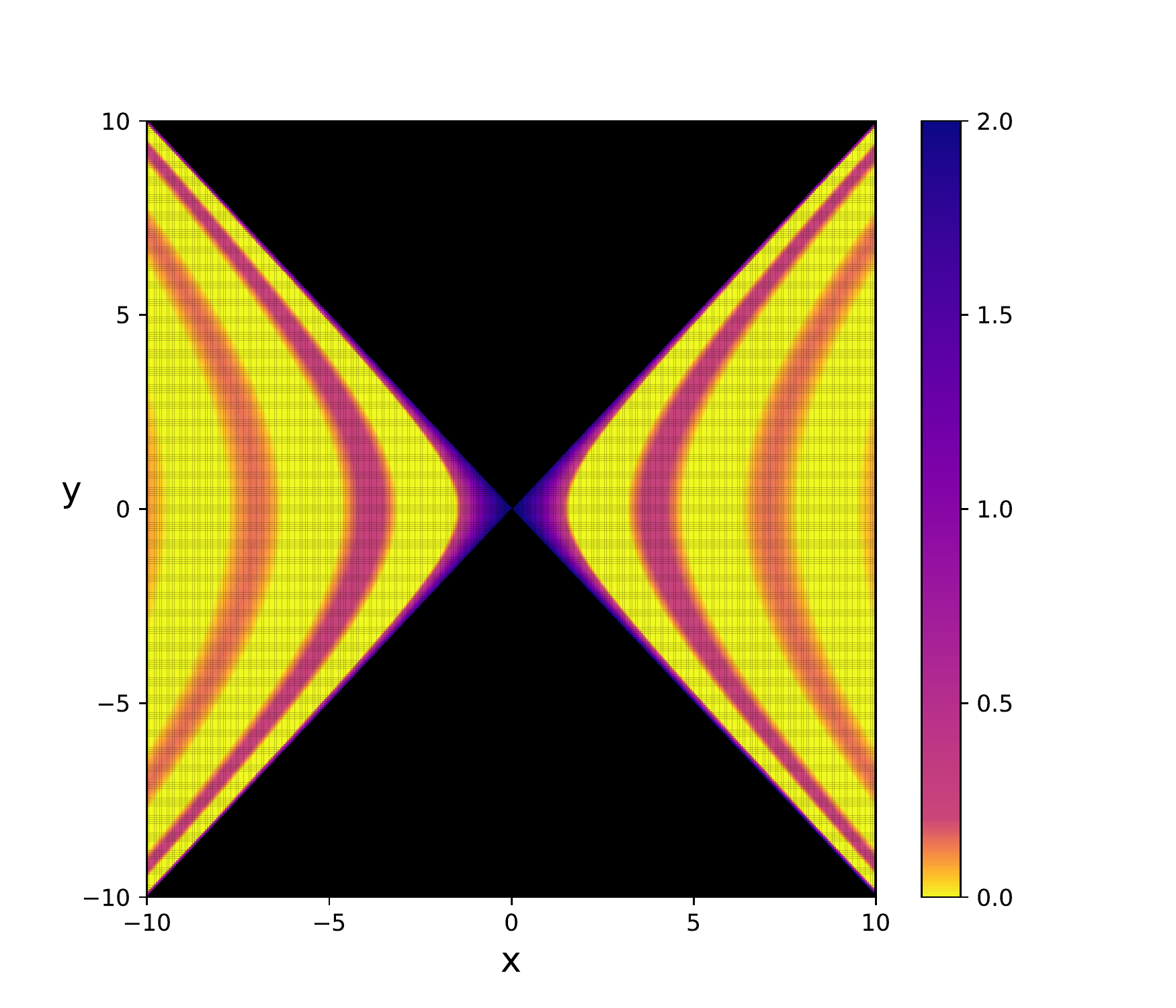}
\caption{(Color online) Anisotropic Friedel oscillations
are plotted from Eq. (\ref{deltarho3}) for positive $E_{\s F}$.
For negative $E_{\s F}$ the plot should be rotated by $90^{\circ}$.}
\label{f3}
\end{figure}

\section{Friedel oscillations}
For a parabolic spectrum, $E_{\bf k}=\frac{\hbar^2k^2}{2m}$,
the Friedel oscillations of the electron density, $\delta n({\bf r})$,
created by a defect, are
isotropic
\begin{equation}
\label{isotropic}
\delta n({\bf r} )\propto
\frac{\sin\left(2k_{\s F}r\right)}
{\left(k_{\s F}r\right)^2},
\end{equation}
where  $k_{\s F}=\left(\frac{2mE_F}{\hbar^2}\right)^{1/2}$
is the Fermi wave vector.

Below we generalize the derivation of  $\delta n({\bf r})$
to the case of a hyperbolic spectrum Eq. (\ref{hyperbolic})
and demonstrate that it assumes the  form

\begin{equation}
\label{deltarho1}
\delta n({\bf r} )\propto
\frac{\sin\left(2k_{\s F}\sqrt{x^2-y^2}\right)}
{\left(k_{\s F} \sqrt{x^2-y^2}  \right)^2}.
\end{equation}
Outside the quadrants $|x|>|y|$ the correction
$\delta\rho({\bf r} )$ falls off exponentially at large $r\gg k_{\s F}^{-1}$.

On the opposite side of the transition, $E_{\s F}<0$,
the result Eq. (\ref{deltarho1}) transforms into

\begin{equation}
\label{deltarho2}
\delta n({\bf r} )\propto
\frac{\sin\left(2k_{\s F}\sqrt{y^2-x^2}\right)}
{\left(k_{\s F} \sqrt{y^2-x^2}  \right)^2},
\end{equation}
with $k_{\s F}=\left(\frac{2m|E_F|}{\hbar^2}\right)^{1/2}$.
Therefore, the crossing from positive to negative
$E_{\s F}$ is accompanied by rotation of the Friedel
oscillations pattern by $90^{\circ}$.
At finite temperature, $T$, the oscillations are
cut off at distance $r_{\s T}$ such that
$k_{\s F}r_{\s T} \sim \frac{E_{\s F}}{T}$, so that
in the transition region $E_{\s F}\sim T$ the oscillations
effectively disappear.

\subsection{Derivation}
Consider a short-range impurity with potential $U({\bf r})$.
It creates the following correction to the free-electron wave functions,
$\Psi_{{\bf k}}({\bf r})$

\begin{equation}
\label{correction}
\delta \Psi_{{\bf k}}({\bf r})    =\sum_{{\bf k}'}\frac{U_{{\bf kk'}}}{E_{{\bf k}}-E_{{\bf k}'}}
\Psi_{{\bf k}'}({\bf r}).
\end{equation}
Then the electron density,
\begin{equation}
\label{density}
n({\bf r})=\sum_{{\bf k}}|\Psi_{{\bf k}}({\bf r})+
\delta \Psi_{{\bf k}}({\bf r})|^2\Theta\left(E_{\s F}-E_{{\bf k}}\right),
\end{equation}
acquires the following correction

\begin{equation}
\label{deltan}
\begin{gathered}
\delta n({\bf r}) = 2\sum_{{\bf k}}\Theta(E_{\s F}-E_{\bf k})\Psi^*_{\bf k}\delta\Psi_{\bf k} \\
=2U_0\sum_{{\bf k},{\bf k}'}\Theta(E_{\s F}-E_{\bf k})\frac{\Psi^*
_{\bf k}\Psi_{{\bf k}'}}{E_{{\bf k}}-E_{{\bf k}'}},
\end{gathered}
\end{equation}
where $U_0 =U_{{\bf kk'}}=\int d{\bf r}U({\bf r})$, and $\Theta(z)$
is a step-function. It is convenient to rewrite Eq. (\ref{deltan}) in the form

\begin{equation}
\label{deltan1}
\delta n({\bf r}) = \frac{U_0}{4\pi^2}\int\limits_{-\infty}^{E_{\s F}}dE_1
\int\limits_{-\infty}^{\infty}dE_2
\frac{\Phi(E_1,{\bf r})\Phi^{\ast}(E_2,{\bf r})}
{E_{1}-E_{2}},
\end{equation}
where we have introduced an auxiliary function
\begin{equation}
\label{Phi}
\begin{gathered}
\Phi(E,{\bf r})=\int d{\bf k}
e^{i{\bf k}{\bf r}}\delta\left(E-E_{\bf k}    \right) \\
=\iint d{ k_{x}} d{ k_{y}}
\delta{\Big[E-\frac{\hbar^2}{2m}\left(k_{x}^2-k_{y}^2\right)\Big]}e^{i{ k_{x}}x+i{ k_{y}}y}.
\end{gathered}
\end{equation}
To establish the analytical form of $\Phi(E,{\bf r})$ we switch to the new variables
\begin{eqnarray}
\label{pxpy}
{ k_{x}}={ p_{x}}\cos\varphi_{\bf {r}}+{ p_{y}}\sin\varphi_{\bf {r}},\nonumber \\
{ k_{y}}={ p_{x}}\sin\varphi_{\bf {r}}-{ p_{y}}\cos\varphi_{\bf {r}},\nonumber \\
\end{eqnarray}
 where $\varphi_{\bf {r}}$ is the azimuthal
angle of ${\bf r}$. Then Eq. (\ref{Phi}) takes the form
\begin{equation}
\label{Phi1}
\begin{gathered}
\Phi(E,{\bf r})= \iint d{ p_{x}} d{ p_{y}} \\
\times\delta{\Big[\frac{2mE}{\hbar^2}-(p_{x}^2-p_{y}^2)
\cos2\varphi_{\bf r}-2p_{x}p_{y}\sin2\varphi_{\bf r}\Big]}e^{ip_{x}{r}}.
\end{gathered}
\end{equation}
Note that $p_y$ is present only in the argument of the $\delta$-function.
To perform the integration over $p_y$ we factorize this argument
\begin{equation}
\label{factorized}
\begin{gathered}
\Phi(E,{\bf r})=\frac{1}{\cos2\varphi_{\bf r}}\int d{ p_{x}}e^{i{ p_{x}}{r}}\int d{ p_{y}} \\
\times \delta \Biggl[
\Bigg(p_{y}+p_{x}\frac{\sin2\varphi_{\bf {r}}}{\cos2\varphi_{\bf {r}}}+\sqrt{\frac{p_{x}^2}{\cos^2 2\varphi_{\bf {r}}}-\left(\frac{2mE}{\hbar^2}\right)\frac{1}{\cos2\varphi_{\bf {r}}}}\Bigg) \\
\times\left(p_{y}+p_{x}\frac{\sin2\varphi_{\bf {r}}}{\cos2\varphi_{\bf {r}}}-\sqrt{\frac{p_{x}^2}{\cos^2 2\varphi_{\bf {r}}}-\left(\frac{2mE}{\hbar^2}\right)\frac{1}{\cos2\varphi_{\bf {r}}}}\right)
\Biggr].
\end{gathered}
\end{equation}
Now the integration over $p_y$ is straightforward and yields
\begin{equation}
\label{Phi2}
\Phi(E,{\bf r})=
\frac{1}{2}\int d{ p_{x}}\frac{e^{i{ p_{x}}{r}}}{\sqrt{p_{x}^2-\left(\frac{2mE}{\hbar^2}\right)\cos2\varphi_{\bf {r}}}}.
\end{equation}
The upper limit in the integral Eq. (\ref{Phi2}) is infinity. The lower limit
depends on the sign of $E\cos2\varphi_{{\bf r}}$. When the sign is negative, there is no
pole in the denominator. Then the lower limit is zero and the integral reduces to the
Macdonald function. For positive  $E\cos2\varphi_{{\bf r}}$ the lower limit is
$\left[\frac{2mE}{\hbar^2\cos2\varphi_{{\bf r}}}\right]^{1/2}$.
The integral then reduces to the Bessel function of the second kind.
Combining both cases, we write

$$
\Phi(E,{\bf r})=\frac{m}{2\pi^2\hbar^2} \Biggl\{
	\begin{array}{ll}
		-Y_0\Big(k_E\sqrt{x^2-y^2}\Big),\quad E(x^2-y^2)>0,\\
		K_0\Big(k_E\sqrt{y^2-x^2}\Big),\quad E(x^2-y^2)<0,
	\end{array}
	\Biggr.
$$
where $k_E=\left(\frac{2mE}{\hbar^2}\right)^{1/2}$.
In the case of a parabolic spectrum the function $\Phi(E,{\bf r})$ is simply the
Bessel function $J_0(k_Er)$.

Now the expression for $\Phi(E,{\bf r})$ should be substituted into Eq. (\ref{deltan1}).
Similarly to the parabolic spectrum, one has to use the large-argument asymptote of
$\Phi(E,{\bf r})$. We see that for $|y|>|x|$ the Macdonald function decays exponentially,
so that there are no oscillations in two quadrants $|y|>|x|$.  For quadrants $|x|>|y|$
the long-distance asymptote of $Y_0(z)$ is $\sin\left(z-\frac{\pi}{4} \right)$, and
differs by a phase $\pi/2$ from the asymptote, $\cos\left(z-\frac{\pi}{4} \right)$, of $J_0(z)$.
Thus, the product $Y_0(z_1)Y_0(z_2)$
contains $\cos\left(z_1+z_2-\frac{\pi}{2}\right)$
in the same way as the product $J_0(z_1)J_0(z_2)$
only with {\em opposite sign}. This allows to proceed
directly to the result for $\delta n({\bf r})$

\begin{equation}
\label{deltarho3}
\delta n({\bf r} )=
\frac{mU_0\sin\left(2k_{\s F}\sqrt{x^2-y^2}\right)}
{2\pi^2\hbar^2\left(x^2-y^2\right)}.
\end{equation}
Eq. (\ref{deltarho3}) illustrates the general connection\cite{BenaReview}
between the Friedel oscillations and the underlying spectrum.

\section{Spectrum renormalization}

We start from the textbook expression\cite{Mahan}
for the exchange self-energy

\begin{equation}
\label{selfenergy}
\Sigma({\bf{k}})=-\int\frac{d^2k'}{(2 \pi)^2}V({\bf{k}}-{\bf{k'}})~n_{\bf{k'}},
\end{equation}
where $n_{\bf{k'}}=\Theta(E_F-E_{{\bf k}'})$ is the Fermi distribution and $V({\bf q})$ is the Fourier component of the electron-electron interaction. We first assume that $E_F=0$
and choose for $V({\bf{q}})$ the screened Coulomb potential
\begin{equation}
\label{vq}
{V}({\bf{q}})=\frac{2\pi e^2}{\epsilon({q}+\kappa)},
\end{equation}
where $\epsilon$ is a bare dielectric constant and $\kappa$ is the inverse screening
radius which we will determine later.

Obviously, the integral over $k'$ diverges at large $k'$ leading to a
general energy shift independent of $k$.
To calculate the spectrum renormalization we subtract this shift
and get

\begin{equation}
\label{withshift}
\Sigma({\bf{k}})-\Sigma(0)=-\frac{2\pi e^2}{\epsilon}\int\frac{d^2k'}{(2\pi)^2}\frac{{k'}-|{\bf{k}}'-{\bf{k}}|}{(k'+\kappa)(|\bf{k}-\bf{k'}|+\kappa)}n_{{\bf{k}}'}.
\end{equation}
Now the integral Eq. (\ref{withshift}) converges at $
k'\sim \kappa$. We are interested in the
spectrum renormalization in the vicinity of the transition.
Assuming that $k \ll \kappa$, we expand
the integrand in parameter $\frac{k}{k'}$.
This yields
\begin{equation}
\label{expanded}
\begin{gathered}
\Sigma({\bf{k}})-\Sigma(0)\\=
-\frac{2 \pi e^2}{\epsilon}
\int\frac{d^2k'}{(2\pi)^2}\Bigg[\frac{\frac{{\bf{k}}
\cdot\bf{k}'}{k'}+
\frac{({\bf{k}}\cdot{\bf{k}}')^2-k^2k'^2}{2k'^3}}
{(k'+\kappa)^2}+\frac{ \left(\frac{{\bf{k}}
\cdot\bf{k}'}{k'}\right)^2  }{(k'+\kappa)^3}\Bigg]n_{{\bf{k}}'}.
\end{gathered}
\end{equation}
The expansion in Eq. (\ref{expanded}) is carried out to the
second order in
$\frac{k}{k'}$, since the first-order
term vanishes upon the angular integration.
Indeed, this term changes sign
upon replacement $\varphi_{{\bf k}'} \rightarrow \left(\pi-\varphi_{{\bf k}'}\right)$,
where $\varphi_{{\bf k}'}$ is the polar angle of the vector ${\bf k}'$.
On the other hand, the argument of $n_{{\bf{k}}'}$ contains $\cos 2\varphi_{{\bf k}'}$,
and it does not change upon this replacement.
Thus the integration of the linear term over $\varphi_{{\bf k}'}$ yields zero.
The second and the third terms in Eq. (\ref{expanded})
give rise to the following
$k^2$-correction to the spectrum

\begin{equation}
\label{angles}
\begin{gathered}
\Sigma({\bf{k}})-\Sigma(0) \\=
\frac{2 \pi e^2}{\epsilon}k^2\int\frac{d^2k' }{(2\pi)^2}
\Bigg[
\frac{\sin^2(\varphi_{\bf{k}}-\varphi_{\bf{k'}})}{2k'(k'+\kappa)^2}
-\frac{\cos^2(\varphi_{\bf{k}}-\varphi_{\bf{k'}})}{(k'+\kappa)^3}         \Bigg]n_{{\bf{k}}'}.
\end{gathered}
\end{equation}
It is instructive to rewrite this correction in the form

\begin{equation}
\label{angles1}
\begin{gathered}
\Sigma({\bf{k}})-\Sigma(0) =\\
-\frac{2 \pi e^2k^2}{\epsilon}\!\!\int\frac{d^2k' }{(2\pi)^2}
\Bigg[\frac{(k'-\kappa)+(3k'+\kappa)\cos2(\varphi_{\bf{k}}-\varphi_{\bf{k'}})}{4k'(k'+\kappa)^3}\Big]n_{{\bf{k}}'}.
\end{gathered}
\end{equation}

We expect that interactions preserve the structure of the spectrum, $k^2\cos2\varphi_{\bf k}$. On the other hand, the first term in the numerator of Eq. (\ref{angles1}) leads
to the isotropic $k^2$-correction. But it is easy to check that the condition
\begin{equation}
\int\limits_0^{\infty}dk' \frac{k'-\kappa}{(k'+\kappa)^3}=0
\end{equation}
is met, so that the coefficient in front of  $k^2$-term is zero.
Final result for the spectrum renormalization reads

\begin{equation}
\label{reproduces1}
\begin{gathered}
\Sigma \left({\bf{k}}\right)-\Sigma(0)= \\
-k^2\cos 2 \varphi_{\bf_{k}}\Big(\frac{e^2}{8\pi \epsilon}\Big)
\int\limits_{0}^{\infty}\frac{dk'(3k'+\kappa)}{(k'+\kappa)^3}
\int\limits_{0}^{2\pi}d\varphi_{\bf_{k'}}n_{\bf_{k'}}\cos2\varphi_{\bf_{k'}}.
\end{gathered}
\end{equation}
For our choice $E_{\s F}=0$
the integrals over $k'$ and over $\varphi_{\bf_{k'}}$
get decoupled. The first integral is equal to $2\kappa^{-1}$, while the second integral is equal to $-2$.
It is convenient to cast Eq. (\ref{reproduces1}) into the form of the renormalized mass in the spectrum Eq. (\ref{hyperbolic})
\begin{equation}
\label{renormalized}
\frac{1}{m_{\s eff}}=\frac{1}{m}\Bigg(1+\frac{2me^2}{\pi\epsilon\kappa \hbar^2}   \Bigg).
\end{equation}

At finite $E_{\s F}$, away from the transition, the dependence on $E_{\s F}$
comes from the integral over $\varphi_{\bf_{k}}$ in Eq. (\ref{reproduces1}). However, the leading dependence on $E_{\s F}$ originates from the parameter $\kappa$.
Within the random-phase approximation the expression for the
inverse screening radius reads
\begin{equation}
\label{kappa}
\kappa=\frac{2\pi e^2}{\epsilon}\nu,
\end{equation}
where $\nu$ is the density of states at the Fermi level.
In fact, $\nu$ diverges in the limit $E_{\s F}\rightarrow 0$. Indeed, on has

\begin{equation}
\label{nu}
\nu(E)=2\int \frac {d^2k}{(2\pi)^2}
\delta \left[E-\frac{\hbar^2}{2m}
\left(k_x^2-k_y^2\right)\right]
=\frac{2m}{\pi^2\hbar^2}
\ln\left(\frac{\tilde E_{\s F}}{|E|}   \right).
\end{equation}
Substituting Eq. (\ref{nu}) into Eq. (\ref{kappa}),
we arrive  to the following expression for renormalized mass
\begin{equation}
\label{renormalized1}
\frac{1}{m_{\s eff}}=\frac{1}{m}\Bigg[1+\frac{1}{2\ln\left({\tilde E}_{\s F}/|E_{\s F}|\right)    }\Bigg].
\end{equation}

\section{Polarization Operator}
Polarization operator for 2D electron gas with a parabolic
spectrum was calculated for the first time by F. Stern.\cite{Stern1967}
Below we calculate polarization operator for a hyperbolic spectrum
Eq. (\ref{hyperbolic}). We start from the definition
\begin{equation}
\label{operator}
\large{\Pi}(\omega,{\bm q})=\sum_{{\bf k}}
\frac{n_{{\bf k}}-n_{{\bf k+q}}}
{\hbar\omega + E_{\bf k}-E_{{\bf k} +{\bf q}}}.
\end{equation}
As a first step,
we cast Eq.~(\ref{operator})
into the form
\begin{equation}
\label{operator1}
\begin{gathered}
\large{\Pi}(\omega,{\bm q})\\
=\sum_{{\bf k}}n_{{\bf k}}\Bigg[\frac{1}{\hbar\omega+E_{{\bf k}}-E_{{\bf k+q}}}-\frac{1} {\hbar\omega+E_{{\bf k-q}}-E_{{\bf k}}}\Bigg].
\end{gathered}
\end{equation}
Introducing, similarly to Eq. (\ref{pxpy}), the new variables
\begin{eqnarray}
\label{pxpy1}
{ k_{x}}={ p_{x}}\cos\varphi_{\bf {q}}+{ p_{y}}\sin\varphi_{\bf {q}},\nonumber \\
{ k_{y}}={ p_{y}}\cos\varphi_{\bf {q}}-{ p_{x}}\sin\varphi_{\bf {q}},\nonumber \\
\end{eqnarray}
and replacing the sum by the integral, we obtain

\begin{equation}
\label{operator2}
\begin{gathered}
\large{\Pi}(\omega,{\bm q})=\\\!\!\iint\frac{d p_x d p_y}{(2\pi)^2}
\Theta\Bigg[ \frac{2mE_{\s F}}{\hbar^2}-\left( p_x^2-p_y^2 \right)\cos2\varphi_{{\bf q}}-2p_x p_y\sin2\varphi_{{\bf q}}   \Bigg]\\
\!\!\times
\Bigg [\frac{1}{\hbar \omega +\frac{\hbar^2q^2}{2m}\cos2\varphi_{{\bf q}}-\frac{\hbar^2}{m}p_xq }   \!-\!\frac{1}{\hbar \omega -\frac{\hbar^2q^2}{2m}\cos2\varphi_{{\bf q}}-\frac{\hbar^2}{m}p_xq }       \Bigg].
\end{gathered}
\end{equation}
Note that the argument of the $\Theta$-function has the same form
as the argument of the $\delta$-function in Eq. (\ref{factorized})
with $\varphi_{{\bf q}}$ instead of $\varphi_{{\bf r}}$.
Then the integration over $p_y$ is straightforward

\begin{equation}
\label{operator3}
\begin{gathered}
\large{\Pi}(\omega,{\bm q})=\int\limits_{-\infty}^{\infty}\frac{dp_x}{2\pi^2}
\Bigg[
\frac{p_x^2}{\cos^2 2\varphi_{{\bf q}}}
-
\frac{2mE_{\s F}}{\hbar^2\cos2\varphi_{{\bf q}}}     \Bigg]^{1/2}\times \\
\Bigg
[\frac{\hbar\omega+\frac{\hbar^2q^2}{2m}\cos2\varphi_{{\bf q}}}
{\left(\hbar\omega+\frac{\hbar^2q^2}{2m}\cos2\varphi_{{\bf q}}\right)^2-\left(\frac{\hbar^2 p_xq}{m}\right)^2}   \\-
\frac{\hbar\omega-\frac{\hbar^2q^2}{2m}\cos2\varphi_{{\bf q}}}
{\left(\hbar\omega-\frac{\hbar^2q^2}{2m}\cos2\varphi_{{\bf q}}\right)^2-\left(\frac{\hbar^2 p_xq}{m}\right)^2}   \Bigg],
\end{gathered}
\end{equation}
where the first square bracket is the result of integration
over $p_y$, and in the second square bracket we have isolated
the parts even in $p_x$.

It is convenient to rewrite Eq. (\ref{operator3}) as follows

\begin{equation}
\label{operator4}
\begin{gathered}
\large{\Pi}(\omega,{\bm q})\\=\frac{m}{\pi^2 \hbar^2q \cos2\varphi_{\bf q}}\int dp_{x}\left[p_{x}^2-A\right]^{1/2}
\left[\frac{\beta_{+}}{\beta_{+}^2-p_{x}^2}-
\frac{\beta_{-}}{\beta_{-}^2-p_{x}^2}\right],
\end{gathered}
\end{equation}
where  the parameters $A$, $\beta_{+}$, and $\beta_{-}$
are defined as

\begin{equation}
\label{A}
\begin{gathered}
A=\frac{2mE_{F}\cos{2\varphi_{\bf q}}}{\hbar^2},\\
\beta_{\pm}=\frac{m}{\hbar^2 q}\left(\hbar\omega\pm\frac{\hbar^2 q^2}{2m}\cos{2\varphi_{\bf q}}\right).
\end{gathered}
\end{equation}

Now the integration in Eq. (\ref{operator4}) can be performed explicitly. The main contribution to the polarization operator comes from log-divergence of the integral at large $p_x$. This divergence is cut off at $p_x \sim \left( 2m{\tilde E}_{\s F}/\hbar^2 \right)^{1/2}$.
The $\omega$ and $q$-dependencies are given by the sub-leading terms

\begin{equation}
\label{operator5}
\begin{gathered}
\large{\Pi}(\omega,{\bm q})=\frac{m}{\pi^2 \hbar^2q \cos{2\varphi_{\bf q}}}\\
\times\left[-(\beta_{+}-\beta_{-})
\ln\left({\frac{\tilde{E_{\s F}}}{ |E_{\s F}\cos{2\varphi_{\bf q}|}}}\right)\right. \\
\left.
-\left(\frac{\beta_{+}^2-A}{\beta_{+}}\right)G\left(\frac{A}{\beta_{+}^2}\right)
+\left(\frac{\beta_{-}^2-A}{\beta_{-}}\right)G\left(\frac{A}{\beta_{-}^2}\right)
\right],
\end{gathered}
\end{equation}
where the function $G(z)$ is defined as
\begin{equation}
\label{FunctionG}
G(z)=\int\frac{d{\cal P}}{({\cal P}^2-z)^{1/2}({\cal P}^2-1)}.
\end{equation}
The upper limit in the integral Eq. (\ref{FunctionG}) is infinity.
The lower limit is $p_y=z^{1/2}$ for positive $z$ and $p_y=0$ for negative $z$. Correspondingly, the form of $G(z)$ is different
for $z>0$ and $z<0$. Namely,

$$
G(z)=\Biggl\{
	\begin{array}{ll}
\label{array}
        \frac{1}{\left(z-1 \right)^{1/2}}
        {\scriptstyle \arcsin}\left(\frac{1}{z^{1/2}}\right),\quad z>0, \\
        \frac{\ln\left(\frac{1}{|z|^{1/2}}+\sqrt{1+\frac{1}{|z|}} \right)}{\left(|z|+1  \right)^{1/2}},\quad z<0,
	\end{array}
	\Biggr.
$$
It is easy to see that $G(z)$ falls off as $1/z$ at large positive $z$ and as $1/|z|$ at large negative $z$.

\subsection{Frequency domain}
Note that the coefficient $\left(\beta_{+}-\beta_{-}\right)$ in front
of leading logarithmic term does not depend on frequency.
In 2D electron gas with parabolic spectrum\cite{Stern1967}
the analog of the combinations $\beta_{\pm}^2-A$ in Eq. (\ref{operator5})
has the form
$\Big[\left(\hbar\omega -\frac{\hbar^2q^2}{2m}\right)^2-
2\frac{\hbar^2q^2{\tilde E}{\s F}}{m} \Big]^{1/2}$.
At small $q$, the polarization operator acquires an imaginary part, which is responsible
for the ac conductivity, when
$\omega>\left(\frac{2{\tilde E}_{\s F}}{m}\right)^{1/2}q$. The corresponding condition
for the hyperbolic spectrum reads $\omega>\left(\frac{2E_{\s F}\cos2\varphi_{\bf q}}{m}\right)^{1/2}q$. Firstly, since ${\tilde E}_{\s F}$, the Fermi energy in the
``bulk", is much bigger than $E_{\s F}$, we conclude that the ac response
at low frequencies is dominated by the proximity to the topological transition.
Secondly, this response is strongly anisotropic.

\subsection{Momentum domain}
In the static limit, $\omega=0$, the polarization operator is a universal function of the dimensionless momentum
\begin{equation}
\label{dimensionless}
Q_{\bf q}=q\left(\frac{\hbar^2}{2mE_{\s F}}\cos2\varphi_{\bf q}   \right)^{1/2}.
\end{equation}
This function has a form

\begin{equation}
\begin{gathered}
\label{staticpolar}
\large{\Pi}({\bf q})=-\frac{m}{2\pi^2\hbar^2} \\
\times\left[\ln\left({\frac{{\tilde E}_{\s F}}{|E_{\s F}\cos2\varphi_{{\bf q}}|}}\right)-
\left(1-\frac{Q_{\bf q}^2}{4}\right)^{1/2}
\frac{\arcsin\left({\frac{Q_{\bf q}}{2}}\right)}{\left(\frac{Q_{\bf q}}{2}\right)}\right],
\end{gathered}
\end{equation}
for $|Q_{\bf q}|<2$. Near the Kohn anomaly the behavior of
$\left(\large{\Pi}({\bf q})-\large{\Pi}(0)\right)$ is singular, $\left(2-|Q_{\bf q}|\right)^{1/2}$. It gives rise to the long-period
Friedel oscillations Eq. (\ref{deltarho3}).

For $|Q_{\bf q}|>2$ the expression for polarization operator reads
\begin{equation}
\label{staticpolar1}
\begin{gathered}
\large{\Pi}({\bf q})
=-\frac{m}{2\pi^2\hbar^2} \times\\
\left\{\ln\left(\!\!{\frac{{\tilde E}}{|E_{\s F}\cos2\varphi_{{\bf q}}|}}\!\!\right)\!\!+
\!\!
\left(1-\frac{4}{Q_{\bf q}^2}\right)^{1/2}
\!\!\!\ln{\left[\left(\frac{Q_{\bf q}}{2}\right)+\!\sqrt{\frac{Q_{\bf q}^2}{4}-1}\right]}\right\}.
\end{gathered}
\end{equation}
As a function of $\left(|Q_{\bf q}|-2\right)$, the behavior of $\large{\Pi}({\bf q})$ is linear.
Note that the behaviors Eqs. (\ref{staticpolar}), (\ref{staticpolar1})
differ from the static polarization operator for the isotropic
spectrum\cite{Stern1967}, where $\large{\Pi}({\bf q})$ is constant for $q<2k_{\s F}$,
while the Kohn anomaly, $\propto \left(q-2k_{\s F}\right)^{1/2}$, is located to the right
from $q=2k_{\s F}$.

To summarize, we have evaluated polarization operator in the entire domain
of frequencies and momenta. In the static limit and for $q_y=0$ our result agrees
with Ref.~\onlinecite{Chi-Ken2016}.
While for parabolic spectrum, $E_{\bf k}=\frac{\hbar^2}{2m}\left(k_x^2+k_y^2\right)$,
the Kohn anomaly corresponds to the condition $q=2k_{\s F}$, the corresponding condition
for the hyperbolic spectrum Eq.~(\ref{hyperbolic}) reads
\begin{equation}
\label{Kohn}
q_x^2-q_y^2=4k_{\s F}^2.
\end{equation}
This condition is illustrated in Fig. \ref{f2}.

\begin{figure}[h!]
\includegraphics[scale=0.65]{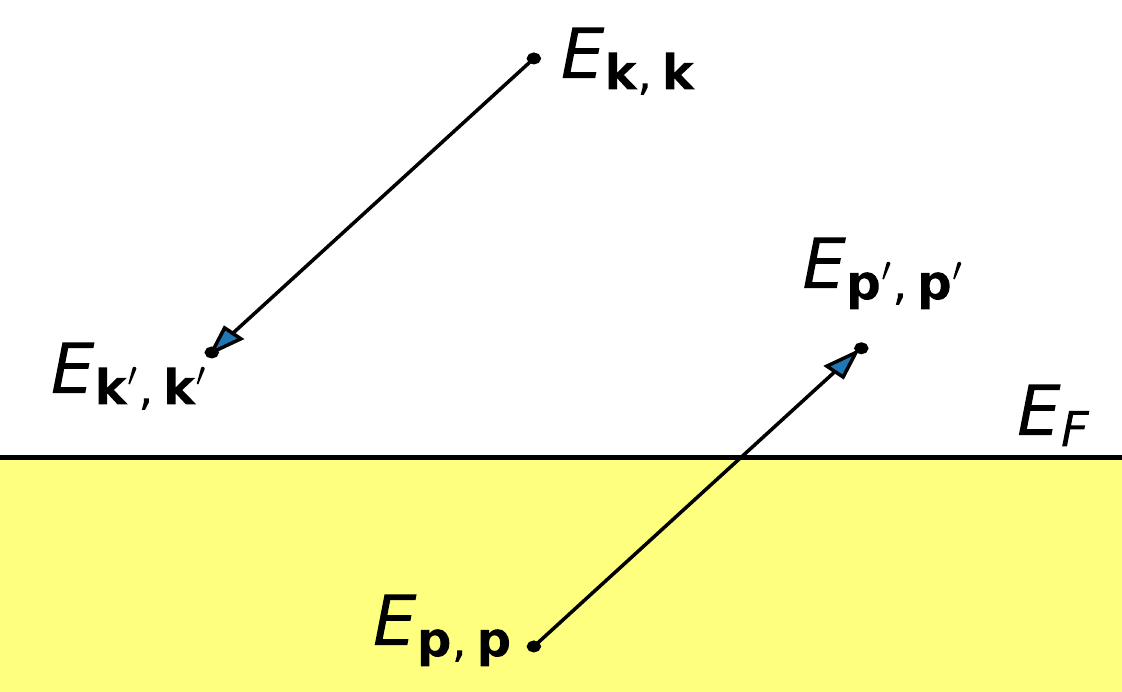}
\caption{(Color online) Illustration of the process responsible for a finite
electron lifetime: Initial electron with momentum ${\bf k}$ and energy $E_{\s k}$ reduces
its energy and goes to the final state $E_{{\bf k}'}$ creating a pair with energies $E_{\bf p}$
and $E_{{\bf p}'}$.}
\label{f4}
\end{figure}

\section{Electron lifetime}

The process which is responsible for a finite lifetime, $\tau_e$,
of an electron with energy, $E_{\bf k}$, above the Fermi level
is creation of an electron-hole pair.
Accurate calculation of $\tau_e$ for an electron gas with a quadratic
spectrum was reported in Refs. \onlinecite{time1}, \onlinecite{time2}.
The result reads

 \begin{equation}
 \label{lifetime}
\frac{1}{\tau_{\bf k}}=\Gamma({\bf k},E_{\bf k})=\frac{E_{\bf k}^2}{4\pi \hbar {\tilde E}_{\s F}}\ln\left(\frac{ {\tilde E}_{\s F}}{E_{\bf k}} \right).
\end{equation}
The $E_{\bf k}^2$-dependence originates from the energy conservation, namely: $E_{\bf k}+E_{\bf p}=E_{{\bf k}'}+E_{{\bf p}'}$, where $E_{{\bf k }'}>{\tilde E}_{\s F}$ is the energy of the secondary electron, while $E_{\bf p}<{\tilde E}_{\s F}$ and  $E_{{\bf p}'}>{\tilde E}_{\s F}$ are the energies of particles constituting an excited pair.
The factor $\ln\left(\frac{ {\tilde E}_{\s F}}{E_{\bf k}} \right)$
originates from the momentum conservation. To generalize Eq. (\ref{lifetime}) to the case of hyperbolic spectrum, we start
from the golden-rule expression for the rate $\tau_{\bf k}^{-1}$

\begin{equation}
\label{lifetime1}
\begin{gathered}
\frac{1}{\tau_{\bf k}}\propto \int\limits_{E_{{\bf k}'}>{\tilde E}_{\s F}}d{\bf k}'
\int\limits_{E_{{\bf p}}<{\tilde E}_{\s F}}d{\bf p}
\int\limits_{E_{{\bf p}'}>{\tilde E}_{\s F}}d{\bf p}'\\
\times \delta \left(E_{\bf k}+E_{\bf p}-E_{{\bf k}'}-E_{{\bf p}'}\right)
\delta\left( {\bf k} +{\bf p}-{\bf k}'-{\bf p}'  \right).
\end{gathered}
\end{equation}
To perform the averaging over the directions of momenta, we introduce
auxiliary variables $E_1$, $E_2$, and $E_3$ and invoke the integral
representation of the $\delta$-function
\begin{equation}
\label{lifetime2}
\begin{gathered}
\frac{1}{\tau_{\bf k}}\propto \int\limits_{{\tilde E}_{\s F}}^{\infty}
dE_1 \int\limits_{-\infty}^{{\tilde E}_{\s F}}dE_2
\int\limits_{{\tilde E}_{\s F}}^{\infty}dE_3~
\delta\left( E_{\bf k}+E_2-E_1-E_3  \right)\\
\times \int d{\bf k}'\int d{\bf p}\int d{\bf p}'
\delta\left(E_{{\bf k}'}-E_1\right) \delta\left(E_{{\bf p}}-E_2\right)    \delta\left(E_{{\bf p}'}-E_3\right)\\
\times \int \frac{d{\bf r}}{(2\pi)^2}\exp\Big[i\left({\bf k}+{\bf p}-{\bf k}'-{\bf p}'\right){\bf r}\Big].
\end{gathered}
\end{equation}
Now the integration over momenta decouples into three integrals of the type $\int d{\bf p}\exp\left(i{\bf pr}\right)\delta\left(E_{\bf p}-E\right)$. For a quadratic spectrum, this integral is expressed through a zero-order Bessel function, $J_0\left( k_Er  \right)$.
Then the integral over ${\bf r}$ in Eq. (\ref{lifetime2})
assumes the form
\begin{equation}
\label{conservation1}
I=\int d{\bf r}  e^{i{\bf k}{\bf r}}J_0(k_{E_{1}}r)J_0(k_{E_{2}}r)J_0(k_{E_{3}}r).
\end{equation}
The angular-averaged $\exp\left( {i{\bf k}{\bf r}}  \right)$
is equal to $J_0(k_Er)$.
The magnitudes of all momenta in Eq. (\ref{conservation1}) are close to
the Fermi momentum, ${\tilde k}_{\s F}$. The long-distance behavior of the product of the
four Bessel functions is $\propto \frac{1}{\left({\tilde k}_{\s F}r\right)^2}$. Then
the integration over $r$ gives rise to the logarithm in Eq.~(\ref{lifetime1}), while
$\frac{1}{{\tilde k}_{\s F}^2}$ generates ${\tilde E}_{\s F}$ in the denominator.

For
a hyperbolic spectrum, the  integral
$\int d{\bf p}\exp\left(i{\bf pr}\right)\delta\left(E_{\bf p}-E\right)$
is given by the function $\Phi(E,{\bf r})$
defined by Eq. (\ref{Phi}).
Then, in place of integral  Eq. (\ref{conservation1}), one has
\begin{equation}
\label{conservation2}
I=\int d{\bf r}  e^{i{\bf k}{\bf r}}\Phi(E_1,{\bf r})\Phi(E_2,{\bf r})\Phi(E_3,{\bf r}).
\end{equation}
Depending on the polar angle of ${\bf r}$, the function $\Phi(E,{\bf r})$ either oscillates with $r$ (in the domain
$-\frac{\pi}{4}<\varphi_{\bf r}<\frac{\pi}{4}$)
or decays with $r$  (in the domain
$\frac{\pi}{4}<\varphi_{\bf r}<\frac{3\pi}{4}$).
In the first domain,  with energies $E_1$, $E_2$, $E_3$ close to
$E_{\s F}$ and wave vector $k$ close to $\left(2mE_{\s F}/\hbar^2\right)^{1/2}$, the slow part of  the integrand in
Eq. (\ref{conservation2}) reproduces, within a numerical factor, the result Eq. (\ref{lifetime}) for the hyperbolic spectrum.

Naturally, the applicability of Eq. (\ref{lifetime}) requires that
$|E_{\bf k}-E_{\s F}| \ll E_{\s F}$. In the vicinity of the topological transition $E_{\s F}$
is small and the estimate for the lifetime follows from Eq. (\ref{lifetime}) upon setting $E_{\s F} \sim E_{\bf k}$.
We conclude that, in the vicinity of the topological transition, $\frac{\hbar}{\tau_e} \sim E$.

Assume now that the Fermi level is $E_{\s F}=0$.
The question of interest is how $\tau_e$ depends on the direction, $\varphi_{\bf k}$, of the momentum of the initial electron.
For $E_{\s F}=0$, energy conservation requires that, when the energy $E_1$ is positive, the energy $E_2$ is negative while the energy $E_3$ is positive, see Fig. \ref{f4}. Then Eq. (\ref{conservation2}) assumes the form

\begin{equation}
\label{conservation3}
\begin{gathered}
I=\int\limits_0^{\infty}dr r\int\limits_0^{2\pi}d\varphi_{\bf r}   \exp\Big[{ikr\cos(\varphi_{\bf k}-\varphi_{\bf r})}\Big]\\
\times Y_0(k_1r\cos2\varphi_{\bf r})K_0(k_2r\cos2\varphi_{\bf r})Y_0(k_3r\cos2\varphi_{\bf r}),
\end{gathered}
\end{equation}
where $k$ is the magnitude of momentum of initial electron, $k_1$ and $k_3$ are momenta of the secondary electrons,
and $-k_2$ is the momentum of a hole.

To find the dependence of $\tau_e$ on $\varphi_{{\bf k}}$ we
introduce instead of $r$ a new variable $z=kr\cos2\varphi_{\bf r}$ and
obtain

\begin{equation}
\label{conservation3}
\begin{gathered}
I=\frac{1}{k^2}\int\limits_0^{2\pi}\frac{d\varphi_{\bf r}}{\cos^22\varphi_{\bf r}} \int\limits_0^{\infty}dz z  \exp\Bigg[iz\frac{\cos\left(\varphi_{\bf k}-\varphi_{\bf r}\right)}{\cos2\varphi_{\bf r}}\Bigg]\\
\times Y_0\left(\frac{k_1}{k}z\right)K_0\left(\frac{k_2}{k}z\right)Y_0\left(\frac{k_3}{k}z\right).
\end{gathered}
\end{equation}
The  form Eq. (\ref {conservation3}) suggests that the
main contribution to the integral comes from the vicinity of $\varphi_{\bf r}~\approx~\pm\frac{\pi}{4}, \pm\frac{3\pi}{4}$.
For these $\varphi_{\bf r}$ the exponent in the integrand rapidly
oscillates with $z$. The exceptions are the vicinities of
$\varphi_{\bf k}=\pm\frac{\pi}{4}, \pm\frac{3\pi}{4}$ when
$\cos\left(\varphi_{\bf k}-\varphi_{\bf r}\right)$ turns to zero
when  $\varphi_{\bf r}$ is close to $\pm\frac{\pi}{4}, \pm\frac{3\pi}{4}$.

As an example, consider a situation $\varphi_{\bf k} \approx -\frac{\pi}{4}$ and set $\varphi_{\bf r} =\frac{\pi}{4}+\psi_{\bf r}$,
with $\psi_{\bf r} \ll 1$. The exponent in Eq.~(\ref {conservation3})
does not oscillate for $\psi_{\bf r} > \psi_{\s min}$, where
$\psi_{\s min} =|\cos\left(\varphi_{\bf k}-\frac{\pi}{4} \right)|$.
Then the angular integration in Eq. (\ref {conservation3}) yields

\begin{equation}
\int\limits_{\psi_{\s min}}^{\infty}\frac{d\psi_{\bf r}}{\psi_{\bf r}^2}
=\frac{1}{\psi_{\s min}}=\frac{1}{|\cos\left(\varphi_{\bf k}-\frac{\pi}{4} \right)|}.
\end{equation}
The above result suggests that the lifetime, $\tau_e$, shortens dramatically for certain directions of momentum of an electron.
Physical explanation of such a shortening  is that the cost of creation of a pair by electron with these directions of momentum is anomalously low.

%

\section{Concluding Remarks}

\noindent({\em i}) Ballistic correction to the conductivity\cite{Gold1986,Zala,Adamov},
$\sigma\left(E_{\s F},T\right)$,
of a 2D electron gas has the form  $ \frac{\delta \sigma}{\sigma}  \sim \lambda\left(\frac{T}{E_F}\right)$, where $\lambda$ is the interaction
parameter.\cite{Zala} The origin
of this correction is electron scattering from the potential created by Friedel
oscillations surrounding individual impurities. The amplitude of this process is sharply peaked
at the scattering angle $\pi$. For this angle, the momentum transfer is close to $2k_{\s F}$, the wave vector of the Friedel oscillation.
For the hyperbolic spectrum, while the wave vector of the Friedel oscillations depends on the direction, but the mechanism of Refs. \onlinecite{Gold1986}, \onlinecite{Zala}
still  applies. It gets modified as illustrated in Fig. \ref{f2}. Backscattering takes place between disjoint parts of the Fermi surface. Smallness of $E_{\s F}$ makes the ballistic correction progressively pronounced in the
vicinity of the transition.

\noindent({\em ii}) While calculating the spectrum renormalization we assumed that
the form of interaction is screened Coulomb, see  Eq. (\ref {vq}).
In fact, the static polarization operator Eq. (\ref {staticpolar}) contains a sub-leading
term, $\left(1-\frac{Q_{\bf q}^2}{4}\right)^{1/2}$ describing the Kohn anomaly. This term is strongly anisotropic. An interesting question is how this anisotropy
affects the spectrum renormalization. Denote with $\delta \kappa({\bf q})$ the correction to the inverse screening radius, describing the Kohn
anomaly, $\delta \kappa ({\bf q}) \propto \left(1-\frac{\hbar^2q^2}{8mE_{\s F}}
\cos2\varphi_{{\bf q}}\right)^{1/2}$.
Expanding the interaction $V({\bf q})$ with respect to $\delta \kappa ({\bf q})$, we get
\begin{equation}
\label{vq1}
\delta V({\bf{q}})=-\frac{2\pi e^2\delta \kappa ({\bf q}) }{\epsilon({q}+\kappa)^2}.
\end{equation}
This correction to $V({\bf{q}})$ gives rise to the following correction to the self-energy
\begin{equation}
\label{deltasigma}
\delta\Sigma({\bf k}) \propto \int d{\bf k}' \delta \kappa ({\bf k}-{\bf k}')
\Theta\left(E_{\s F}-\frac{\hbar^2k'^2}{2m}\cos2\varphi_{{\bf k}'}\right).
\end{equation}
At small momenta, $|k|\ll k_{\s F}$, Eq. (\ref{deltasigma}) leads to the following
contribution to the spectrum renormalization
\begin{equation}
\label{deltasigma1}
\delta\Sigma({\bf k}) \propto E_{\bf k}\int d{\bf k}'
\frac{E_{{\bf k}'}}{\left(E_{\s F}-E_{{\bf k}'}\right)^{3/2}}\Theta\left(E_{\s F}-E_{{\bf k}'}   \right).
\end{equation}
This correction diverges, the divergence comes from the vicinity of
 $\varphi_{\bf k}=\pm \frac{\pi}{4}, \pm \frac{3\pi}{4}$.

\noindent ({\em iii}) Divergence of lifetime for directions of momenta close to $\varphi_{{\bf q}} =\pm \frac{\pi}{4}, \pm \frac{3\pi}{4}$ also hints at strong renormalization of the spectrum for these momenta.

\noindent ({\em iv}) With regard to observables, interaction-induced modification of the effective mass manifests itself in the magneto-oscillations. Behavior of magneto-oscillations in the vicinity of the topological transition constitutes a subfield called the magnetic breakdown,
see e.g. the review Ref. \onlinecite{breakdown}.
As the Fermi level is swept through the topological transition, the
period of magneto-oscillations doubles.
The width of the domain of $E_{\s F}$ where this doubling takes place is $\sim \frac{\hbar^2}{ml^2}$,
where $l$ is the magnetic length. For $E_{\s F} \gg \frac{\hbar^2}{ml^2} $ the coupling of the
semiclassical trajectories is determined by tunneling under the magnetic barrier.\cite{barrier}
Then the dependence of the effective mass on $E_{\s F}$
affects the barrier transmission.

Electron lifetime is measured in 2D-2D tunneling experiments.\cite{Tunnel1,Tunnel2} The lifetime defines the width of the
peak in the tunnel conductance measured versus the dc bias applied between
the layers.

\noindent({\em v}) There is a conceptual similarity between Friedel oscillations of
elections of the electron density created by an impurity and the oscillations of the
spin density created by a magnetic impurity\cite{Roth1966}. In this regard, long-period Friedel oscillations
in the vicinity of the topological transition are similar to the long-period behavior
of the RKKY interaction established in Ref.  \onlinecite{Golosov1991}.


\section{Acknowledgements}

\vspace{2mm}
The work was supported by the Department of Energy,
Office of Basic Energy Sciences, Grant No.  DE-FG02-06ER46313.


\begin{thebibliography}{30}


\bibitem{Lifshitz1960} I. M. Lifshitz, ``Anomalies of electron characteristics of a metal in the high pressure region," Sov. Phys. JETP {\bf 11}, 1130 (1960).


\bibitem{Experiment1983} V. S. Egorov and A. N. Fedorov,
``Thermopower of lithium-magnesium alloys at the $2~1/2$ -order transition,"
Sov. Phys. JETP {\bf 58}, 959 (1983).

\bibitem{Experiment1984} N. V. Zavaritskii and I. M. Suslov,
``Structural features in the thermopower of a two-dimensional electron gas near
topological transitions," Sov. Phys. JETP {\bf 60}, 1243 (1984).


\bibitem{1}
E. A. Yelland, J. M. Barraclough, W. Wang,
K.~ V.~ Kamenev, and A. D. Huxley,
  ``High-field superconductivity at an electronic topological transition in
URhGe," Nat.
Phys. {\bf 7}, 890 (2011).

\bibitem{2}
M. Orlita, P. Neugebauer, C. Faugeras, A. L. Barra, M.~Potemski, F. M. D. Pellegrino, and D. M. Basko,
``Cyclotron Motion in the Vicinity of a Lifshitz Transition in Graphite,"
Phys. Rev. Lett. {\bf 108},  017602 (2012).

\bibitem{3} A. Varleta, M. Mucha-Kruczy{\'n}ski, D. Bischoff, P. Simonet, T. Taniguchi, K. Watanabe, V. Fal'ko, T. Ihn, and K. Ensslin,
``Tunable Fermi surface topology and Lifshitz transition in bilayer graphene,"
Synthetic Metals {\bf 210},  19 (2015).

\bibitem{4}
H.-R. Chang, J. Zhou, H. Zhang, and Y. Yao,
``Probing the topological phase transition via density oscillations in silicene and
germanene," Phys. Rev. B {\bf 89}, 201411(R) (2015).


\bibitem{5}
H. Chi, C. Zhang, G. Gu, D. E Kharzeev,
Xi Dai, and Qiang Li,
``Lifshitz transition mediated electronic transport
anomaly in bulk ZrTe$_5$," New J. Phys. {\bf 19},
015005 (2017).

\bibitem{6} M. E. Barber, A. S. Gibbs, Y. Maeno, A. P. Mackenzie, and C. W. Hicks,
``Resistivity in the Vicinity of a van Hove Singularity: Sr$_2$RuO$_4$ under Uniaxial Pressure",
Phys.  Rev. Lett. {\bf 120}, 076602 (2018).


\bibitem{7}
P. Di Pietro, M. Mitrano, S. Caramazza, F. Capitani, S. Lupi, P. Postorino, F. Ripanti, B. Joseph, N. Ehlen,
A. Gr{\"u}neis, A. Sanna, G. Profeta, P. Dore, and A. Perucchi1, ``Emergent Dirac carriers across a pressure-induced Lifshitz transition in black phosphorus," Phys. Rev. B {\bf 98}, 165111 (2018).

\bibitem{8}
T. Nishimura, H. Sakai, H. Mori, K. Akiba, H. Usui, M. Ochi, K. Kuroki, A. Miyake, M. Tokunaga, Y. Uwatoko, K. Katayama, H. Murakawa, and N. Hanasaki, ``Large Enhancement of Thermoelectric Efficiency Due to a Pressure-Induced Lifshitz Transition in SnSe,"
Phys. Rev. Lett. {\bf 122}, 226601 (2019).


\bibitem{Varlamov1985} A. A. Varlamov and A. V. Pantsulaya,``Anomalous kinetic properties of metals near the Lifshitz topological transition,"
Sov. Phys. JETP {\bf 62}, 1263 (1985).


\bibitem{Varlamov1986} A. A. Varlamov and A. V. Pantsulaya, ``Absorption of iongitudinal sound in metals near the Lifshitz topologlcai transition,"
Sov. Phys. JETP {\bf 64},  1319 (1986).


\bibitem{Blanter1990} Ya. M. Blanter, A. A. Varlamov, and A. V. Pantsulaya,
``Giant oscillations of the magnetothermoelectric power of a metal near an electronic topological transition," Sov. Phys.
JETP {\bf 70}, 695 (1990).

\bibitem{Blanter1991}
Ya. M. Blanter, A. A. Varlamov, and A. V. Pantsulaya, ``Thermoelectric power and topological transitions in quasi-2D electron systems," Sov. Phys.  JETP {\bf 73}, 688 (1991).

\bibitem{Ablyazov1991} N. N. Ablyazov, M. Yu. Kuchiev, and M. E. Raikh,
``Topological transition and its connection with the conductivity
and thermopower anomalies in two-dimensional systems,"
Phys. Rev. B {\bf 44}, 8802 (1991).

\bibitem{Golosov1991} D. I. Golosov and M. I. Kaganov, ``Spatial correlation of conduction electrons near the electron-topological
transition in a metal,"  Sov. Phys. JETP {\bf 74}, 186 (1992).


\bibitem{Mobius} M. I. Kaganov and A. M{\"o}bius,
``Effect of Fermi-liquid interaction on a phase transition of order $2~1/2$,"
Sov. Phys. JETP {\bf 59}, 405 (1984).


\bibitem{VarlamovReview} Y. M. Blanter, M. I. Kaganov, A. V. Pantsulaya, and A. A. Varlamov,
``The theory of electronic topological transitions,"
Phys. Rep. {\bf 245},  159 (1994).









\bibitem{Weyl}
K. Seo, C. Zhang, and S. Tewari,
``Thermodynamic signatures for topological phase transitions to Majorana and Weyl superfluids in ultracold Fermi gases."
Phys. Rev. A {\bf 87}, 063618 (2013).

\bibitem{Thorus} J.-W. Rhim and Y. B. Kim,
``Anisotropic density fluctuations, plasmons, and Friedel oscillations
in nodal line semimetal,"   New J. Phys. {\bf 18}, 043010 (2016).

\bibitem{Chi-Ken2016}  C.-K. Lu,  ``Friedel oscillation near a van Hove singularity
in two-dimensional Dirac materials,"
J. Phys.: Condens. Matter {\bf 28}, 065001 (2016).

\bibitem{Graphene}
T. Farajollahpour, S. Khamouei, S. S. Shateri, and A. Phirouznia,
``Anisotropic Friedel oscillations in graphene-like materials:
The Dirac point approximation in wave-number dependent quantities revisited,"
Sci. Rep. {\bf 8}, 2667 (2018).


\bibitem{Volovik} G. E. Volovik, ``Exotic Lifshitz transitions in topological materials,"
Phys. Usp. {\bf 61}, 89 (2018).


\bibitem{Galperin}
Y. M. Galperin, D. Grassano, V. P. Gusynin, A. V. Kavokin, O. Pulci, S. G. Sharapov, V. O. Shubnyi, and A. A. Varlamov, ``Entropy Signatures of Topological Phase Transitions,"
J. Exp. Theor. Phys. {\bf 127}, 958 (2018).

\bibitem{BenaReview}
C. Bena, ``Friedel oscillations: Decoding the hidden physics,"
C. R. Physique {\bf 17}, 302 (2016).

\bibitem{Mahan} G. Mahan, ``Many-Particle Physics,"
Physics of Solids and Liquids (Springer, New York, 2010).


\bibitem{Stern1967} F. Stern, ``Polarizability of a Two-Dimensional Electron Gas,"
Phys. Rev. Lett. {\bf 18}, 546 (1967).

\bibitem{time1} T. Jungwirth and A. H. MacDonald,
``Electron-electron interactions and two-dimensional$-$two-dimensional tunneling,"
Phys. Rev. B {\bf 53}, 7403 (1996).


\bibitem{time2}
L. Zheng and S. Das Sarma,
``Coulomb scattering lifetime of a
two-dimensional electron gas,"
Phys. Rev. B {\bf 53}, 9964 (1996).


\bibitem{Gold1986}
A.  Gold  and  V. T.  Dolgopolov,
``Temperature dependence of the conductivity for the
two-dimensional electron gas: Analytical results for low temperatures,"
Phys.  Rev.  B {\bf 33},  1076 (1986).


\bibitem{Zala} G. Zala,  B. N. Narozhny, and I. L. Aleiner,
``Interaction corrections at intermediate temperatures:
Longitudinal conductivity and kinetic equation,"
Phys. Rev. B {\bf 64}, 214204 (2001).

\bibitem{Adamov}
Y. Adamov, I. V. Gornyi, and A. D. Mirlin,
``Interaction effects on magneto-oscillations in a two-dimensional electron gas,"
Phys. Rev. B {\bf 73}, 045426 (2006).


\bibitem{breakdown} M. I. Kaganov and A. A. Slutskin, ``Coherent
magnetic breakdown," Phys. Rep. {\bf 98},  189 (1983).

\bibitem{barrier} G. E. Zil'berman, ``Electron in a periodic
electric and homogeneous magnetic field, II", Sov. Phys. JETP {\bf 6}, 299 (1958).


\bibitem{Tunnel1} J. P. Eisenstein, T. J. Gramila, L. N. Pfeiffer, and K. W. West,
``Probing a two-dimensional Fermi surface by tunneling,"
Phys. Rev. B {\bf 44}, 6511 (1991).



\bibitem{Tunnel2} S. Q. Murphy, J. P. Eisenstein, L. N. Pfeiffer, and K. W. West, ``Lifetime of two-dimensional electrons measured by tunneling spectroscopy,"
Phys. Rev. B {\bf 52}, 14825 (1995).













\bibitem{Roth1966}
L. M. Roth, H. J. Zeiger, and T. A. Kaplan,
``Generalization of the Ruderman-Kittel-Kasuya-Yosida
Interaction for Nonspherical Fermi Surfaces," Phys. Rev. {\bf 149}, 519 (1966).
\end{thebibliography}
\end{document}